\documentclass{article}

 \usepackage[preprint]{neurips_2025}

\usepackage[utf8]{inputenc} 
\usepackage[T1]{fontenc}    
\usepackage{hyperref}       
\usepackage{url}            
\usepackage{booktabs}       
\usepackage{amsfonts}       
\usepackage{nicefrac}       
\usepackage{xcolor}         

\usepackage{graphicx}
\usepackage{subfigure}
\usepackage{booktabs, multirow}
\usepackage{kotex}

\usepackage{amsmath} 
\usepackage{amssymb}

\usepackage{colortbl}    
\usepackage[table]{xcolor}  
\definecolor{lightblue}{RGB}{220,230,245}
\title{Instance-Specific Test-Time Training \\for Speech Editing in the Wild}
\workshoptitle{The First Workshop on Generative and Protective AI for Content Creation}

\author{%
  Taewoo~Kim    \\ Korea Electronics \\ Technology Institute \\
  \texttt{kimtaewoo@keti.re.kr} \\
  \And
  Uijong~Lee    \\ Korea Electronics \\ Technology Institute \\
  \texttt{jjong2201@keti.re.kr} \\
  \And
  Hayoung~Park  \\ Korea Electronics \\ Technology Institute \\
  \texttt{hyformal@keti.re.kr} \\
  \AND
  Choongsang~Cho\\ Korea Electronics \\ Technology Institute \\
  \texttt{ideafisher@keti.re.kr} \\
  \And
  Nam~In~Park   \\ National Forensic \\ Service \\
  \texttt{naminpark@korea.kr} \\
  \And
  Young~Han~Lee \\ Korea Electronics \\ Technology Institute \\
  \texttt{yhlee@keti.re.kr} \\
}

\begin{document}

\maketitle

\begin{abstract}
Speech editing systems aim to naturally modify speech content while preserving acoustic consistency and speaker identity. However, previous studies often struggle to adapt to unseen and diverse acoustic conditions, resulting in degraded editing performance in real-world scenarios. To address this, we propose an instance-specific test-time training method for speech editing in the wild. Our approach employs direct supervision from ground-truth acoustic features in unedited regions and indirect supervision in edited regions via auxiliary losses based on duration constraints and phoneme prediction. This strategy mitigates the bandwidth discontinuity problem in speech editing, ensuring smooth acoustic transitions between unedited and edited regions. Additionally, it enables precise control over speech rate by adapting the model to target durations via mask length adjustment during test-time training. Experiments on in-the-wild benchmark datasets demonstrate that our method outperforms existing speech editing systems in both objective and subjective evaluations. 

\end{abstract}

\section{Introduction}
Speech editing is a task that modifies speech content while preserving speaker identity and acoustic characteristics. It plays a pivotal role in speech applications such as disfluency removal, content creation, and speech de-identification. One key application is speech de-identification, which removes or replaces personally identifiable information such as names or credit card numbers, enabling the production of privacy-sensitive content without re-recording. However, achieving seamless integration between the edited and unedited regions remains challenging, particularly under diverse and unpredictable acoustic conditions. For practical deployment, it is essential that models maintain naturalness and speaker-identity consistency, ensure content fidelity, and remain robust to in-the-wild acoustic variability.

Recent advances in speech editing~\cite{tan2021editspeech, bai20223, jiang2023fluentspeech, peng2024voicecraft, le2023voicebox, wang2024speechx, cambara2024mapache} have been largely enabled by the architectures and principles of neural text-to-speech~\cite{yu20c_interspeech,ren2021fastspeech,jeong2024efficient,chen2025neural}. Tan~et~al.~\cite{tan2021editspeech} proposed an autoregressive (AR) model that divides the audio into edited and unedited regions and merges forward and backward generation results to ensure smooth acoustic transitions. Bai~et~al.~\cite{bai20223} introduced a speech editing system that demonstrated robust performance for unseen speakers by leveraging speech-text alignment embeddings. Furthermore, Jiang~et~al.~\cite{jiang2023fluentspeech} adopt a context-aware spectrogram denoiser to achieve high-quality and expressive speech. Although these various approaches have achieved promising results in restricted environments such as audiobooks, their effectiveness in real-world speech scenarios remains largely unexplored.

Compared to controlled studio settings, speech editing in the wild is considerably more challenging, as it involves various complex factors such as background noise, reverberation, and bandwidth mismatch. Peng et al.~\cite{peng2024voicecraft} introduced VoiceCraft, a neural codec language model for speech editing, which improves context representation in an AR model through token rearrangement and refined causal masking. However, it depends on large-scale datasets and lacks fine-grained control over the prosody of the edited regions.

In this work, we propose an instance-specific test-time training (TTT) approach to address the challenges of speech editing in real-world scenarios. Our method fine-tunes a speech editing model for each test sample at inference time, leveraging direct supervision from unedited regions and indirect supervision from edited regions. To this end, we apply TTT in two stages, targeting the duration predictor and the spectrogram denoiser. The duration predictor is optimized using phoneme duration loss on unedited regions and auxiliary duration losses on the edited regions, enhancing prosodic consistency and enabling control over speech rate by adjusting the length of edited segments. The spectrogram denoiser is optimized with reconstruction loss and phoneme classification loss, which mitigates overfitting and improves speaker similarity and acoustic consistency under real-world conditions. Experimental evaluations demonstrate that, despite being pretrained on clean speech data, our method exhibits robust editing performance on acoustically challenging audio samples. 

\section{Related Work}

\subsection{Speech Editing}
Early approaches to speech editing focused on modifying acoustic parameters rather than altering the linguistic content of speech. Traditional signal processing techniques such as PSOLA~\cite{charpentier1986diphone}, MBROLA~\cite{dutoit96_icslp}, and WORLD~\cite{morise2016world} enabled prosody modification, including pitch and duration adjustments, by directly manipulating the waveform. However, their reliance on direct waveform manipulation limited their ability to perform linguistic edits, such as inserting or replacing words. 

Building on advances in automatic speech recognition (ASR) and neural text-to-speech (TTS) systems, research on speech editing has shifted toward detecting the target text segment to be modified and synthesizing replacement speech accordingly. Notable approaches include EditSpeech~\cite{tan2021editspeech}, which employs bidirectional fusion for smooth boundary transitions, and A$^3$T~\cite{bai20223} with alignment-aware acoustic-text pretraining. However, these models still struggle to achieve robustness under diverse real-world conditions. More recently, VoiceCraft~\cite{peng2024voicecraft}, a neural codec-based model, has been proposed to improve robustness in the wild, but it still lacks fine-grained controllability over prosody and duration in the edited regions. In this work, we explore methods to enhance robustness in real-world scenarios while enabling controllable prosody, without relying on large-scale speech datasets.

\subsection{Test-Time Training for Speech Editing}

Test-time training (TTT)~\cite{sun2020test} is a paradigm in which a model is adapted to each test instance during inference, typically by optimizing a self-supervised or auxiliary loss on the given input~\cite{sun2020test}. This allows the model to leverage instance-specific information, improving generalization to distribution shifts without retraining on large datasets. TTT has been successfully applied for domain adaptation~\cite{wangtent, osowiechi2023tttflow} and speech processing tasks such as speech recognition~\cite{kim23f_interspeech} and speech enhancement~\cite{kim2021test}. 

In the context of speech editing, TTT remains largely unexplored. The capability to fine-tune a speech editing model on each test utterance could mitigate mismatches between training and deployment conditions, particularly under diverse noise, reverberation, or bandwidth constraints. Our work extends this idea by introducing an instance-specific TTT framework that applies direct supervision on unedited regions and auxiliary constraints on edited regions, enabling prosodic control and improved acoustic consistency in real-world speech editing scenarios.

\section{Method}
\label{sec:method}
\begin{figure*}[t]
\centerline{\includegraphics[width=\textwidth]{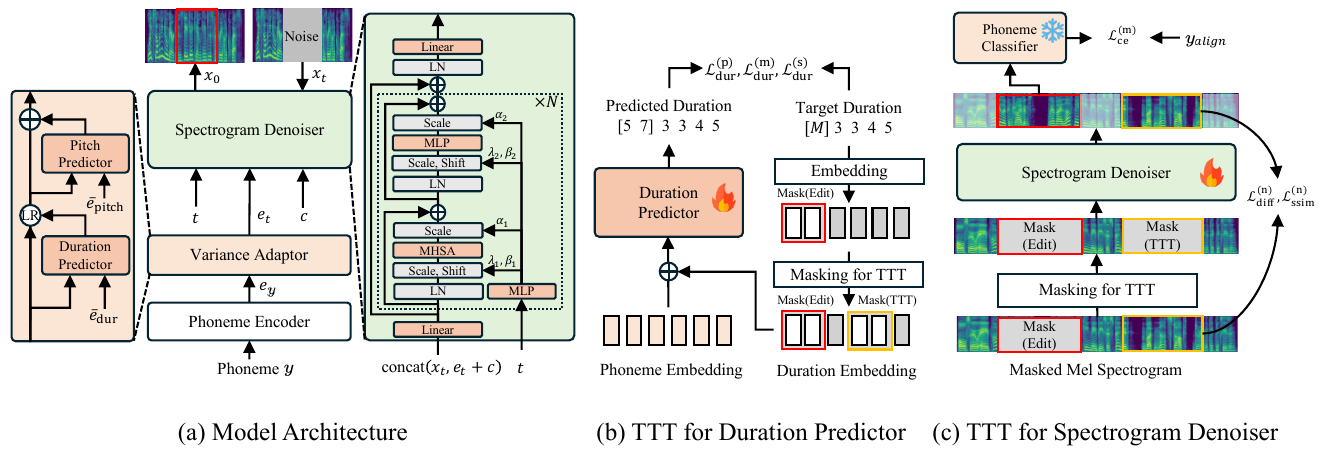}}
\caption{Overview of the proposed framework. In subfigure (a), LR denotes the length regulator. In subfigure (b), $M$ indicates the length of the edit mask, which is required for test-time training (TTT) of the duration predictor. In subfigures (b) and (c), ``Masking for TTT'' refers to randomly masking unedited regions to compute the reconstruction loss during TTT. Red boxes indicate edit regions, and yellow boxes represent randomly masked regions for TTT. The flame icon denotes modules that are updated during TTT, whereas the snowflake icon indicates modules whose parameters remain frozen.}
\label{fig1}
\end{figure*}

In this section, we present our proposed method, which consists of three components: model architecture, train-time training, and test-time training. The overall framework is illustrated in Fig.~\ref{fig1}. We first present a backbone model for speech editing and its train-time training procedure, followed by a detailed description of our test-time training strategy. Each component is discussed in the following subsections.

\subsection{Model Architecture}
Our backbone for speech editing is built upon the architecture of FluentSpeech \cite{jiang2023fluentspeech}, with a key modification: we replace the non-causal WaveNet \cite{vandenoord16_ssw} used in the spectrogram denoiser with a Diffusion Transformer (DiT) \cite{peebles2023scalable} to enhance context modeling and generation performance. The overall architecture of the model consists of a phoneme encoder, a variance adaptor, and a spectrogram denoiser, as illustrated in Fig.~\ref{fig1}(a). The phoneme encoder converts a phoneme sequence $y \in \mathbb{Z}^{1 \times N}$, where $N$ is the length of the phoneme sequence, into $D$-dimensional phoneme representations $e_y \in \mathbb{R}^{D \times N}$. The variance adaptor, which includes a duration predictor and a pitch predictor, predicts the duration and pitch of the masked regions to transform $e_y$ into aligned hidden representations $e_t \in \mathbb{R}^{D \times T}$, where $T$ is the target length of the output sequence. In this process, both the duration predictor and the pitch predictor take $e_y$ along with the masked contextual representations, $e_{dur}$ and $e_{pitch}$, as inputs. Finally, the spectrogram denoiser takes as input the aligned hidden representation \( e_t \), the noisy mel-spectrogram \( x_t \) at the diffusion timestep \( t \), and the condition \( c \), which consists of the speaker embedding and the masked mel-spectrogram embedding. It then predicts the clean target mel-spectrogram \( x_0 \)~\cite{jiang2023fluentspeech, huang2022prodiff} by performing the reverse process of the generator-based diffusion model, formulated as \( f_\theta(x_t, c, e_t, t) \).

\subsection{Train-Time Training}
During training, following \cite{jiang2023fluentspeech}, the model is trained with reconstruction losses on duration, pitch, and mel-spectrogram prediction. First, the duration and pitch losses are computed using L2 loss as follows:
\begin{align}
    \mathcal{L}_{\mathrm{dur}} &= \|d- f_{\mathrm{dp}}(e_{y}, \bar{e}_{\mathrm{dur}}) \|_2^2, \\
    \mathcal{L}_{\mathrm{pitch}} &= \|p- f_{\mathrm{pp}}(e_{t}, \bar{e}_{\mathrm{pitch}}) \|_2^2, 
\end{align}
where $f_{\mathrm{dp}}$ and $f_{\mathrm{pp}}$ represent the duration predictor and the pitch predictor, respectively, and $p$ and $d$ are the target pitch and duration in the masked regions. $\bar{e}_{\mathrm{dur}}$ and $\bar{e}_{\mathrm{pitch}}$ denote masked embeddings provided to each predictor. This encourages the predictors to infer prosodic patterns directly from corrupted or incomplete contextual cues.
For mel-spectrogram loss, the output of the spectrogram denoiser is computed against the ground truth mel-spectrogram using both L1 loss and structural similarity index (SSIM) loss \cite{ren2022revisiting}:
\begin{align}
    \mathcal{L}_{\mathrm{diff}} &= \left\| f_{\theta}(x_t, c, e_t, t) - x_0 \right\|_1 , \\
    \mathcal{L}_{\mathrm{ssim}} &= 1-\operatorname{SSIM}\left( f_{\theta}(x_t, c, e_t, t), x_0 \right), 
\end{align}
where $x_0$ denotes the ground-truth mel-spectrogram at the masked regions, and $x_t$ is the noisy mel-spectrogram at timestep $t$, obtained through the forward diffusion process as formulated in~\cite{jiang2023fluentspeech}. 

Finally, the overall training objective is formulated as a weighted sum of the above components:
\begin{align}
    \mathcal{L}_{\mathrm{train}} 
    &= \lambda_{\mathrm{dur}} \mathcal{L}_{\mathrm{dur}} 
     + \lambda_{\mathrm{pitch}} \mathcal{L}_{\mathrm{pitch}} 
     + \lambda_{\mathrm{diff}} \mathcal{L}_{\mathrm{diff}} 
     + \lambda_{\mathrm{ssim}} \mathcal{L}_{\mathrm{ssim}},
\end{align}
where $\lambda_{\mathrm{dur}}, \lambda_{\mathrm{pitch}}, \lambda_{\mathrm{diff}},$ and $\lambda_{\mathrm{ssim}}$ are coefficients that control the relative contributions of each loss term. This formulation ensures that the model jointly optimizes prosodic characteristics and spectral fidelity.

\subsection{Test-Time Training}
We propose a test-time training (TTT) strategy to enhance prosodic and acoustic consistency at inference time. This approach follows a commonly used instance-level TTT scheme \cite{sun2020test}, in which the model is adapted individually for each test sample. Our method consists of two stages that fine-tune the duration predictor and spectrogram denoiser.

\subsubsection{TTT for Duration Predictor}
In the first stage, TTT is applied to the duration predictor, a key module that predicts the durations within the edited region by capturing the prosodic context from the input text and surrounding unedited regions. To adapt to variations in speaking style across different test conditions, the duration predictor is fine-tuned at test time. To facilitate TTT, we apply additional random masking to the duration embeddings outside the edited region, as illustrated in Fig.~\ref{fig1}(b). For each test sample, multiple input variants are created using different random masking patterns. These variants are grouped into a batch, increasing the batch size and enabling the model to adapt using a diverse set of masked inputs derived from the test sample. The model is then fine-tuned using a phoneme-level duration loss, denoted as $\mathcal{L}_{\mathrm{dur}}^{\mathrm{(p)}}$, computed at these masked positions. A mask-level duration loss, $\mathcal{L}_{\mathrm{dur}}^{\mathrm{(m)}}$, is defined with respect to the sum of the predicted phoneme-level durations within the masked region. A sentence duration loss, $\mathcal{L}_{\mathrm{dur}}^{\mathrm{(s)}}$, is also introduced, based on the total predicted duration of the entire utterance. We define the total TTT loss for the duration predictor as a weighted combination of three L2 losses, where $\lambda_\mathrm{p}$, $\lambda_\mathrm{m}$, $\lambda_\mathrm{s}$ are the weights for phoneme-level, mask-level, and sentence-level duration losses, respectively:
\begin{equation}
\mathcal{L}_{\mathrm{test}}^{\mathrm{DP}} =
\lambda_{\mathrm{p}} \mathcal{L}_{\mathrm{dur}}^{\mathrm{(p)}} +
\lambda_{\mathrm{m}} \mathcal{L}_{\mathrm{dur}}^{\mathrm{(m)}} +
\lambda_{\mathrm{s}} \mathcal{L}_{\mathrm{dur}}^{\mathrm{(s)}} .
\end{equation}

\subsubsection{TTT for Spectrogram Denoiser}
In the second stage, TTT is applied to the spectrogram denoiser to enhance the naturalness and acoustic consistency of the generated speech. Similarly to the previous stage, an additional masking strategy is applied to regions of the mel-spectrogram outside the inference mask, as illustrated in Fig.~\ref{fig1}(c). The spectrogram denoiser is fine-tuned by computing reconstruction losses over these newly masked regions. To maintain intelligibility, we employ a pretrained phoneme classifier to the predicted mel-spectrogram within the inference mask, computing a cross-entropy loss against the aligned phoneme sequence. The total TTT loss for the spectrogram denoiser is defined as a weighted sum of the following terms:
\begin{equation}
\mathcal{L}_{\mathrm{test}}^{\mathrm{SD}}=
\lambda_{\mathrm{diff}} \mathcal{L}_{\mathrm{diff}}^{\mathrm{(n)}} +
\lambda_{\mathrm{ssim}} \mathcal{L}_{\mathrm{ssim}}^{\mathrm{(n)}} +
\lambda_{\mathrm{ce}} \mathcal{L}_{\mathrm{ce}}^{\mathrm{(m)}},
\end{equation}

where the superscripts \( \mathrm{(n)} \) and \( \mathrm{(m)} \) indicate the newly masked region used for reconstruction loss and the inference mask region used for phoneme classification, respectively. This joint optimization encourages the model to produce outputs that are both acoustically consistent and intelligible.

\section{Experiments}
\label{sec:exp}
\subsection{Dataset and Preprocessing}
We use the LibriTTS dataset~\cite{zen19_interspeech}, a multi-speaker English corpus containing approximately 585 hours of speech recorded at 24 kHz. For training on clean speech, we use only the \texttt{train-clean-100} and \texttt{train-clean-360} subsets, totaling about 245 hours from 1,151 speakers. We evaluated the model in both clean and in-the-wild conditions. The clean condition uses the \texttt{test-clean} subset of LibriTTS, while the in-the-wild condition is evaluated using the GigaSpeech test set~\cite{chen21o_interspeech}, which consists of 16 kHz audio recordings from podcasts and YouTube videos. All audio is resampled to 22.05 kHz with 16-bit quantization. 

In our evaluation setup, we randomly sample 400 utterances from each test set for objective evaluation, and 40 utterances for subjective evaluation. To align audio with transcripts, we use the Montreal Forced Aligner (MFA)~\cite{mcauliffe17_interspeech}. For waveform synthesis from mel-spectrograms, we adopt the pretrained UNIVERSAL V1 HiFi-GAN vocoder\footnote{\texttt{\url{https://github.com/jik876/hifi-gan}}}~\cite{kong2020hifi}, which uses a 1024-point fast Fourier transform (FFT), a 256-sample hop size, a 1024-sample window length, and 80 mel-filterbanks covering the frequency range from 0 to 8 kHz. In addition, pitch contours are extracted using Parselmouth\footnote{\texttt{\url{https://github.com/YannickJadoul/Parselmouth}}}.

\subsection{Experimental Setup}
Our proposed speech editing model consists of three main components: a phoneme encoder, a variance adaptor, and a spectrogram denoiser. The spectrogram denoiser adopts Diffusion Transformer (DiT) blocks with zero-initialized adaptive Layer Normalization, configured with 12 Transformer layers, 6 attention heads, and a hidden dimension of 384. All other settings follow \cite{jiang2023fluentspeech}. The total number of parameters is approximately 44M, increasing to 46M when including the phoneme classifier used for test-time training (TTT). The phoneme classifier follows the architecture introduced by \cite{zhang2022visinger}, comprising two feed-forward Transformer blocks \cite{ren2021fastspeech} followed by a linear projection layer. It is trained jointly with the baseline model, but optimized separately using cross-entropy loss to predict the aligned phoneme sequence. More detailed descriptions of the model configuration are provided in Appendix \ref{sec:appendix_a}.

For training, we use 8 diffusion steps ($T=8$), a batch size of 32, and the Adam optimizer with a learning rate of $2\times10^{-4}$. The model is trained for 700K iterations on a single NVIDIA A40 GPU. Each TTT stage consists of 200 fine-tuning steps with the same batch size and optimizer. Only the duration predictor or the spectrogram denoiser is updated during TTT, with all other components remaining frozen. The learning rates for the masked duration predictor and the spectrogram denoiser are set to $2\times10^{-4}$ and $5\times10^{-5}$, respectively. We apply 80\% phoneme-level masking during both training and TTT. During TTT, however, masking is applied only to unedited regions. We set $\lambda_{\mathrm{p}}$, $\lambda_{\mathrm{m}}$, and $\lambda_{\mathrm{s}}$ all to 1.0, and $\lambda_{\mathrm{dur}}$, $\lambda_{\mathrm{pitch}}$, $\lambda_{\mathrm{diff}}$, $\lambda_{\mathrm{ssim}}$, and $\lambda_{\mathrm{ce}}$ to 1.0, 1.0, 0.5, 0.5, and 1.0, respectively.

\subsection{Baselines}
To evaluate the effectiveness of our proposed method, we compare it against two baseline systems: FluentSpeech \cite{jiang2023fluentspeech} and VoiceCraft \cite{peng2024voicecraft}. FluentSpeech is a non-autoregressive framework that incorporates a variance adaptor and a spectrogram denoiser for speech editing. Trained on the same LibriTTS clean subsets as our model, we directly use its official implementation\footnote{\texttt{\url{https://github.com/Zain-Jiang/Speech-Editing-Toolkit}}}. In contrast, VoiceCraft adopts a large-scale autoregressive transformer architecture for speech editing. It is pretrained with 830M parameters on the GigaSpeech XL corpus~\cite{chen21o_interspeech}, which contains approximately 10,000 hours of diverse audio data from podcasts and YouTube videos. We employ the released official model\footnote{\texttt{\url{https://github.com/jasonppy/VoiceCraft}}} for evaluation, representing a large-scale pretraining baseline that contrasts with our framework trained solely on LibriTTS clean subsets.

\subsection{Evaluation Metrics}
\label{sec:evaluation-metrics}

For evaluation, we follow the setup of \cite{bai20223}, where the middle third of each evaluation utterance is masked and reconstructed using the original transcript, allowing for comparison between the ground-truth and the reconstructed audio. Objective metrics include word error rate (WER)\footnote{\texttt{\url{https://huggingface.co/facebook/hubert-large-ls960-ft}}}~\cite{hsu2021hubert}, speaker similarity (SIM)\footnote{\texttt{\url{https://github.com/microsoft/UniSpeech/tree/main/downstreams/speaker_verification}}}~\cite{chen2022wavlm}, and mel-cepstral distortion~(MCD)~\cite{kubichek1993mel}. For subjective evaluation, we use mean opinion score (MOS) and comparative MOS~(CMOS)~\cite{loizou2011speech}, while a detailed description of the evaluation protocol and results is provided in Appendix~\ref{sec:appendix_b}.

\section{Results}
\label{sec:result}
\begin{table}[t]
\centering
\small
\caption{Speech editing performance on LibriTTS and GigaSpeech test sets. ``Proposed'' uses test-time training (TTT). ``DP'' and ``SD'' denote Duration Predictor and Spectrogram Denoiser, respectively.}
\label{table:performance}
\begin{tabular}{lcccccc}
\toprule
\textbf{System} & \textbf{Data (hr)} & \textbf{Params} & \textbf{WER} \(\downarrow\) & \textbf{SIM} \(\uparrow\) & \textbf{MCD} \(\downarrow\) & \textbf{MOS}{\tiny\(\pm\)CI} \(\uparrow\) \\
\midrule
\rowcolor{lightblue}
\multicolumn{7}{c}{\textbf{Test Set: LibriTTS clean}} \\
\midrule
FluentSpeech & LibriTTS(245) & 24M & 4.35 & 0.765 & 4.38 & 3.94{\tiny\(\pm0.04\)} \\
VoiceCraft & GigaSpeech(10,000) & 830M & 5.22 & 0.778 & 4.39 & 3.99{\tiny\(\pm0.04\)} \\
\midrule
Proposed & LibriTTS(245) & 46M & \textbf{4.13} & \textbf{0.815} & \underline{4.02} & \textbf{4.02}{\tiny\(\pm0.04\)} \\
w/o TTT for DP & LibriTTS(245) & 46M & 4.21 & \underline{0.811} & 4.22 & 4.00{\tiny\(\pm0.03\)} \\
w/o TTT for SD & LibriTTS(245) & 44M & \underline{4.20} & 0.789 & \textbf{4.01} & \underline{4.02}{\tiny\(\pm0.04\)} \\
w/o TTT for Both & LibriTTS(245) & 44M & 4.24 & 0.792 & 4.26 & 4.01{\tiny\(\pm0.03\)} \\
\midrule
Ground Truth & - & - & 4.24 & - & - & 4.11{\tiny\(\pm0.03\)} \\
\midrule
\rowcolor{lightblue}
\multicolumn{7}{c}{\textbf{Test Set: GigaSpeech}} \\
\midrule
FluentSpeech & LibriTTS(245) & 24M & 17.88 & 0.662 & 6.16 & 3.69{\tiny\(\pm0.04\)} \\
VoiceCraft & GigaSpeech(10,000) & 830M & 20.72 & \textbf{0.758} & 6.13 & 3.86{\tiny\(\pm0.04\)} \\
\midrule
Proposed & LibriTTS(245) & 46M & \underline{16.88} & \underline{0.725} & \textbf{5.60} & \textbf{3.87}{\tiny\(\pm0.04\)} \\
w/o TTT for DP & LibriTTS(245) & 46M & \textbf{16.85} & 0.711 & 5.92 & \underline{3.86}{\tiny\(\pm0.04\)} \\
w/o TTT for SD & LibriTTS(245) & 44M & 17.46 & 0.687 & \underline{5.64} & 3.79{\tiny\(\pm0.04\)} \\
w/o TTT for Both & LibriTTS(245) & 44M & 17.15 & 0.688 & 5.98 & 3.75{\tiny\(\pm0.04\)} \\
\midrule
Ground Truth & - & - & 16.78 & - & - & 3.88{\tiny\(\pm0.04\)} \\
\bottomrule
\end{tabular}
\end{table}

\begin{table}[t]
\centering
\small
\caption{Ablation study of test-time training components in the Duration Predictor on the GigaSpeech test set.}
\label{table:ablation_ttt_dp}
\begin{tabular}{lcccc}
\toprule
\textbf{Method} & \textbf{WER} \(\downarrow\) & \textbf{SIM} \(\uparrow\)& \textbf{MCD} \(\downarrow\)& \textbf{CMOS} \(\uparrow\)\\
\midrule
TTT for Duration Predictor & \textbf{17.46} & \textbf{0.687} & \textbf{5.64} & \textbf{0} \\
\midrule
w/o Mask Duration Loss (MDL) & 17.49 & 0.681 & 5.80 & $-$0.31 \\
w/o MDL and Sentence Duration Loss & 17.91 & 0.683& 6.08 & $-$0.34 \\
\bottomrule
\end{tabular}
\end{table}

\begin{table}[t]
\centering
\small
\caption{Ablation study of test-time training components in the Spectrogram Denoiser on the GigaSpeech test set.}
\label{table:ablation_ttt_sd}
\begin{tabular}{p{5.0cm} c c c c}
\toprule
\textbf{Method} & \textbf{WER} \(\downarrow\) & \textbf{SIM} \(\uparrow\)& \textbf{MCD} \(\downarrow\)& \textbf{CMOS} \(\uparrow\)\\
\midrule
TTT for Spectrogram Denoiser & \textbf{16.85} & \textbf{0.711} & \textbf{5.92}  & \textbf{0}  \\
\midrule
Replacing CE Loss with CTC Loss & 18.76 & 0.696 & 5.98 & $-$0.54\\
w/o Phoneme Classifier & 20.77 & 0.704 & 5.97 & $-$0.59 \\

\bottomrule
\end{tabular}
\end{table}

\subsection{Performance Evaluation}

Table~\ref{table:performance} summarizes the performance of our proposed method, the baselines, and ablated variants across both LibriTTS and GigaSpeech test sets. We report results using the objective and subjective metrics described in Section~\ref{sec:evaluation-metrics}. 

On the LibriTTS clean test set, our model achieves the best overall performance compared to the baselines across all metrics. It records the lowest WER (4.13) and MCD (4.02), while attaining the highest SIM (0.815) and MOS (4.02), outperforming both FluentSpeech and VoiceCraft despite being trained only on 245 hours of clean data. This demonstrates the effectiveness of our framework under clean conditions.

On the more challenging GigaSpeech test set, our method consistently surpasses FluentSpeech across all metrics and remains competitive with VoiceCraft, which benefits from large-scale pretraining on 10,000 hours of data. Specifically, our model achieves lower WER (16.88 vs. 17.88) and MCD (5.60 vs. 6.13), as well as higher MOS (3.87 vs. 3.69) compared to FluentSpeech. Compared to VoiceCraft, our system performs better in WER, MCD, and MOS, with only a slightly lower SIM (0.725 vs. 0.758). These results highlight the competitiveness of our approach even without large-scale pretraining. To support subjective evaluation, corresponding audio samples are available online\footnote{\texttt{\url{https://rlataewoo.github.io/ttt-editor}}}. 

To confirm the effect of test-time training (TTT), we conducted ablation studies by removing TTT from the duration predictor (DP), the spectrogram denoiser (SD), or both modules. Disabling TTT for the duration predictor increases MCD, likely due to unnatural speech rhythm that in turn degrades spectral quality. Removing TTT for SD leads to reduced speaker similarity, indicating that TTT enhances acoustic coherence. When TTT is removed entirely, performance drops across all metrics, underscoring the importance of test-time adaptation for robust speech editing in the wild. 

\subsection{Ablation Studies on Test-Time Training Components}
We conduct ablation experiments on the GigaSpeech test set to examine the contribution of individual components within our instance-specific test-time training (TTT) framework, focusing on the duration predictor and the spectrogram denoiser.  
\subsubsection{Duration Predictor} 
Table~\ref{table:ablation_ttt_dp} presents the results of TTT applied to the duration predictor. With TTT enabled for the duration predictor, the system achieves the best overall performance within this ablation setting, striking a consistent balance among WER, SIM, MCD, and CMOS. Removing mask-level duration loss (MDL) leads to only marginal changes in WER and SIM, but noticeably increases MCD (5.64~$\rightarrow$~5.80) and decreases CMOS (0~$\rightarrow$~$-$0.31), suggesting a decline in spectral fidelity and perceived naturalness. Further removing sentence-level duration loss results in additional degradation in WER (17.49~$\rightarrow$~17.91) and MCD (5.80~$\rightarrow$~6.08), indicating that sentence-level control helps stabilize temporal alignment at a global level. These findings highlight the complementary effects of both loss terms in enhancing prosodic stability during TTT.

\subsubsection{Spectrogram Denoiser} 
As shown in Table~\ref{table:ablation_ttt_sd}, we further investigate the contributions of the spectrogram denoiser components in the TTT framework. Replacing the cross-entropy (CE) loss with connectionist temporal classification (CTC) loss \cite{graves2006connectionist} leads to considerable degradation in WER (16.85~$\rightarrow$~18.76), SIM (0.711~$\rightarrow$~0.696), and CMOS (0~$\rightarrow$~$-$0.54). Furthermore, removing the phoneme classification branch entirely results in even worse performance, with WER increasing to 20.77 and CMOS dropping to $-$0.59. These results highlight the importance of phoneme classification for intelligibility and acoustic consistency in real-world conditions.

\subsection{Visualizations}
\begin{figure}[t]
\centerline{\includegraphics[width=0.8\columnwidth]{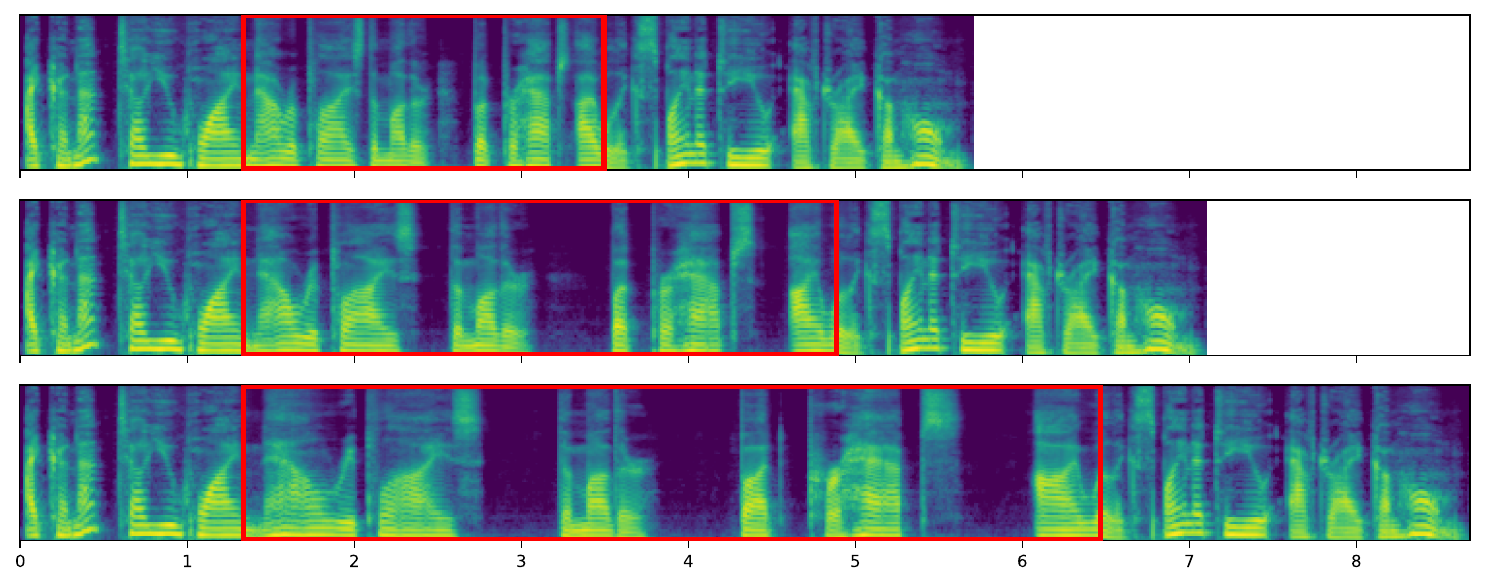}}
\caption{Mel-spectrograms of generated speech at different speech rates using TTT for the duration predictor. The middle row represents the original sentence duration, while the top and bottom rows show $-$20\% and $+$20\% adjustments, respectively. Red boxes indicate the edited regions.}
\label{fig2}
\end{figure}

To qualitatively analyze the effect of test-time training, we provide spectrogram visualizations. Figure~\ref{fig2} demonstrates the controllability of speech rate achieved by applying TTT to the duration predictor. The middle row corresponds to the original sentence duration, while the top and bottom rows represent $-20\%$ and $+20\%$ adjustments, respectively. We observe that TTT enables clear and consistent modifications of speech tempo, without requiring additional duration control modules as in prior approaches~\cite{kim2024masked, effendi2022duration}. These results confirm that TTT provides effective instance-specific prosodic control.  

\begin{figure}
\centerline{\includegraphics[width=0.9\columnwidth]{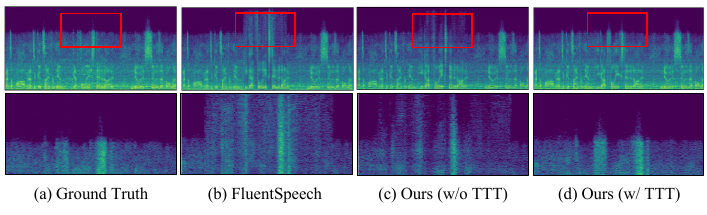}}
\caption{Linear spectrograms of ground-truth and generated speech from different systems. The top panel shows the full spectrogram, while the bottom panel highlights the corresponding regions with red boxes.}
\label{fig3}
\end{figure}

To further assess the effect of TTT, Fig.~\ref{fig3} illustrates its application to the spectrogram denoiser. In real-world scenarios, bandwidth mismatches between edited and unedited regions often result in perceptual discontinuities. Our method alleviates this issue by adapting the model to the acoustic conditions of the input, leading to smoother transitions and enhanced spectral coherence. As shown in Fig.~\ref{fig3}(b), FluentSpeech exhibits prominent bandwidth mismatches at the edited regions, producing audible discontinuities. Ours without TTT (Fig.~\ref{fig3}(c)) partially mitigates these artifacts, benefiting from the DiT-based denoiser which captures longer-range acoustic context compared to the WaveNet architecture used in FluentSpeech. Finally, with TTT applied (Fig.~\ref{fig3}(d)), our model further adapts to input-specific conditions, resulting in the smoothest transitions and the most coherent spectral structure across the edited and unedited regions.

\section{Conclusion}
In this work, we introduce an instance-specific test-time training framework for speech editing under real-world conditions. Our method leverages ground-truth supervision from unedited regions together with auxiliary objectives in edited regions, enhancing both prosodic stability and acoustic consistency. The framework also enables fine-grained control of speech rate through duration adaptation, without requiring explicit duration control modules. Experiments on LibriTTS and GigaSpeech demonstrate that our approach consistently outperforms prior systems across objective and subjective metrics. These results highlight the effectiveness of test-time training for speech editing and demonstrate that, even with limited training data, our framework generalizes well to unseen and diverse conditions, indicating strong potential for real-world adoption.

\section*{Acknowledgments}
This work was supported in part by the Institute of Information and Communications Technology Planning and Evaluation (IITP) Grant funded by Korea Government (MSIT) under Grant RS-2025-02215393 and No.2022-0-00963).

\bibliographystyle{unsrt}
\bibliography{main}


\newpage

\appendix

\section{Model Configuration}
\label{sec:appendix_a}
\begin{table}[h]
\centering
\caption{Hyperparameters of the proposed model.}
\label{tab:model_configuration}
\begin{tabular}{l|l|c}
\toprule
Layer & Hyperparameter & Setting \\
\midrule
\multirow{5}{*}{Text Encoder} 
  & Phoneme Embedding           & 192  \\
  & Encoder Layers              & 4    \\
  & Encoder Hidden              & 192  \\
  & Encoder Heads               & 2    \\
  & Encoder Conv1D Kernel       & 5    \\
  & Encoder Conv1D Filter Size  & 768  \\
  & Encoder Dropout             & 0.0  \\
\midrule
\multirow{4}{*}{Duration Predictor}         
  & Predictor Conv1D Kernel      & 5  \\
  & Predictor Layers   & 3   \\
  & Predictor Conv1D Filter Size & 192   \\
  & Predictor Dropout            & 0.2   \\
\midrule
\multirow{4}{*}{Pitch Predictor}         
  & Predictor Conv1D Kernel      & 5     \\
  & Predictor Layers             & 5     \\
  & Predictor Conv1D Filter Size & 192   \\
  & Predictor Dropout            & 0.2   \\
\midrule
Mel Encoder 
  & Encoder Hidden               & 192 \\
\midrule
\multirow{5}{*}{Spectrogram Denoiser}     
  & Diffusion Embedding          & 384  \\
  & DiT Blocks                   & 12   \\
  & Denoiser Hidden              & 384  \\
  & Denoiser Heads               & 6    \\
  & Denoiser MLP Hidden          & 1536 \\
  & Denoiser Dropout             & 0.1  \\
\midrule
\multirow{6}{*}{Phoneme Classifier}     
  & Classifier Layers             & 2    \\
  & Classifier Hidden             & 256  \\
  & Classifier Heads                        & 2 \\ 
  & Classifier Conv1D Kernel                  & 3 \\
  & Classifier Conv1D Filter Size           & 1024  \\
  & Classifier Dropout                      & 0.5 \\
\midrule
\multicolumn{2}{c|}{\textbf{Total Number of Parameters}} & 45.9M \\
\bottomrule
\end{tabular}
\end{table}

\section{Subjective Evaluation}
\label{sec:appendix_b}

\begin{figure*}[!ht]
    \centering
    \small
    \subfigure[Mean opinion score (MOS) instruction.]{
        \centering
        \includegraphics[width=\linewidth]{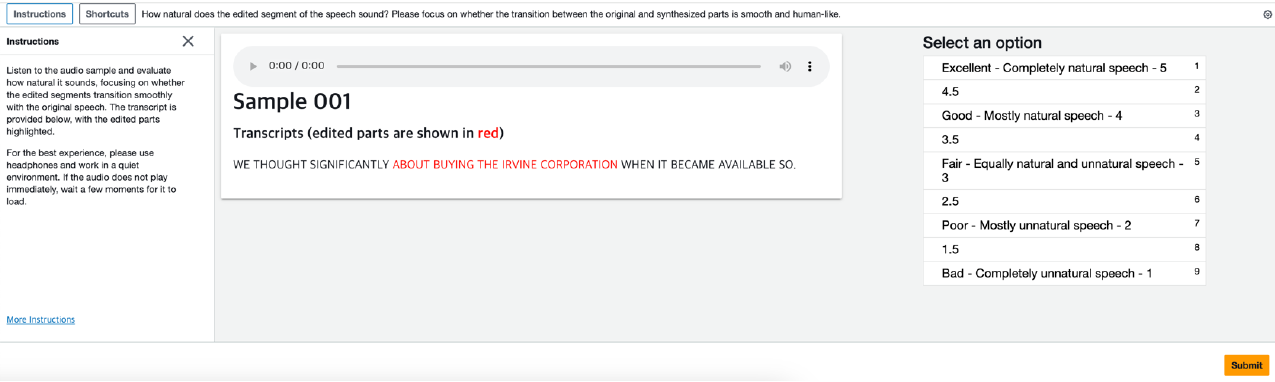}
        \label{fig4_1}
    }
    \subfigure[Comparative mean opinion score (CMOS) instruction.]{
        \centering
        \includegraphics[width=\linewidth]{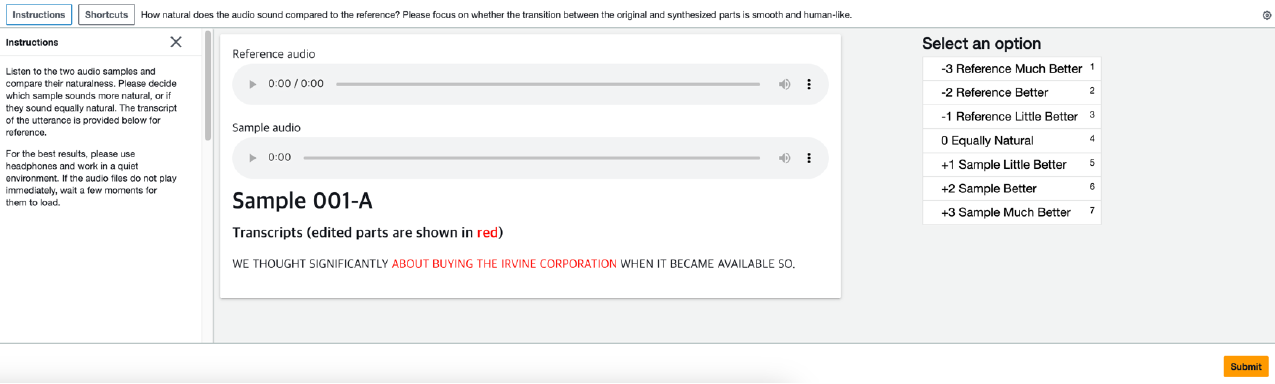}
        \label{fig4_2}
    }
\caption{Instruction interfaces for subjective evaluation tasks.}
\label{fig4}
\end{figure*}
We conducted subjective evaluation using Amazon Mechanical Turk (MTurk)~\footnote{\texttt{\url{https://www.mturk.com/}}}, in terms of both mean opinion score (MOS) and comparative MOS (CMOS). All audio samples were resampled to 22.05 kHz for evaluation. We recruited 40 U.S.-based crowd workers for each test, ensuring that each participant was instructed to submit ratings independently in a quiet environment using headphones. For the MOS test, participants rated the naturalness of each audio sample on a five-point Likert scale with 0.5 increments ranging from 1 (completely unnatural) to 5 (completely natural). As illustrated in Figure~\ref{fig4}(a), the transcript was provided alongside the audio, with edited segments highlighted in red, and workers were instructed to pay particular attention to the smoothness of transitions between original and synthesized segments. For the CMOS test, workers were presented with a pair of utterances: a reference audio and a corresponding sample audio, and they were asked to compare the naturalness of the two, focusing on the seamlessness of transitions between original and synthesized parts on a scale ranging from $-3$ (reference much better) to $+3$ (sample much better), with 0 indicating equal naturalness, as shown in Figure~\ref{fig4}(b).

\section{Limitations and Future Work}
\label{sec:appendix_c}

Despite the promising results, our framework has several limitations that warrant further investigation. First, test-time training (TTT) introduces additional computational overhead, which obstructs real-time deployment in latency-sensitive scenarios. Moreover, while our method demonstrates robustness across diverse conditions, performance degradation may still occur under extreme noise, highly divergent accents, or when only very limited unedited regions are available for supervision. Finally, controllability in our system is primarily restricted to speech rate, leaving other prosodic factors less explored.

Future work will focus on addressing these challenges. One promising direction is the development of parameter-efficient or meta-learning-based TTT approaches that enable faster adaptation suitable for real-time applications. Extending controllability beyond duration to aspects such as pitch, energy, and style is another important avenue. Furthermore, exploring cross-lingual and low-resource scenarios will broaden the applicability of our framework. 

\section{Broader Impacts}
\label{sec:appendix_d}

The proposed test-time training framework for speech editing has the potential to substantially impact both research and real-world applications. By enhancing prosodic stability and acoustic consistency in edited speech, it enables more natural and controllable audio synthesis for content creation, personalized media production, and accessibility technologies. Moreover, its ability to adapt to unseen conditions without large-scale retraining makes it suitable for deployment in diverse scenarios, including low-resource settings where data collection is challenging.

On the other hand, the technology carries risks of misuse, particularly in creating deceptive or harmful audio such as deepfakes. To mitigate these concerns, it is essential to develop safeguards including provenance tracking, watermarking, and detection systems, and to communicate transparently about the framework’s capabilities and limitations. Overall, this work highlights both the societal benefits of more robust and controllable speech editing and the importance of proactive measures to ensure its ethical and responsible deployment.

\end{document}